\documentclass[runningheads]{llncs}
\usepackage{graphicx}
\usepackage{bm}
\usepackage{siunitx}
\usepackage{multirow}
\usepackage{hyperref}
\hypersetup{
	colorlinks=true,
	linkcolor=blue,
	filecolor=blue,      
	urlcolor=blue,
	citecolor=blue,
}

\begin{document}
	\title{Domain-Adversarial Learning for Multi-Centre, Multi-Vendor, and Multi-Disease Cardiac MR Image Segmentation}
	\titlerunning{Domain-Adversarial Cardiac MR Segmentation}
	% \orcidID{0000-1111-2222-3333}
	\author{Cian M. Scannell\inst{1}\orcidID{0000-0001-9240-793X} \and
		Amedeo Chiribiri\inst{1,3} \and
		Mitko Veta\inst{2,3}}
	\authorrunning{C.M Scannell et al.}
	\institute{School of Biomedical Engineering and Imaging Sciences, King's College London, London, UK\\ \and
		Department of of Biomedical Engineering, Eindhoven University of Technology, Eindhoven,
		the Netherlands\\  \and
		Joint last authors \\
		\email{cian.scannell@kcl.ac.uk}}
	\maketitle              % typeset the header of the contribution
	\begin{abstract}
		Cine cardiac magnetic resonance (CMR) has become the gold standard for the non-invasive evaluation of cardiac function. In particular, it allows the accurate quantification of functional parameters including the chamber volumes and ejection fraction. Deep learning has shown the potential to automate the requisite cardiac structure segmentation. However, the lack of robustness of deep learning models has hindered their widespread clinical adoption. Due to differences in the data characteristics, neural networks trained on data from a specific scanner are not guaranteed to generalise well to data acquired at a different centre or with a different scanner. In this work, we propose a principled solution to the problem of this domain shift. Domain-adversarial learning is used to train a domain-invariant 2D U-Net using labelled and unlabelled data. This approach is evaluated on both seen and unseen domains from the M\&Ms challenge dataset and the domain-adversarial approach shows improved performance as compared to standard training. Additionally, we show that the domain information cannot be recovered from the learned features.
	\end{abstract}
	\section{Introduction}
	
	The characterisation of cardiac structure and function is an important step in the diagnosis and management of patients with suspected cardiovascular disease. Cardiac magnetic resonance (CMR) is the method of choice for the non-invasive assessment of cardiac function and allows the accurate quantification of structural and functional parameters such as the chamber volumes and ejection fraction. A limiting factor for the routine analysis of these parameters in the clinic is that it requires the tedious manual delineation of the anatomical structures.
	
	Deep learning has become the state-of-the-art approach for image segmentation and convolutional neural network (CNN)-based approaches have shown huge potential for the automated analysis of cardiac magnetic resonance (CMR) images \cite{Bernard2018,Bai2018,Tao2019,Scannell2020}. Ruijsink et al. \cite{Ruijsink2020} recently showed that it is further possible to automatically compute a wide range of advanced indices of cardiac function, including shape, motion, and strain, from short-axis cine CMR images. This is an important step as it makes it feasible to analyse large numbers of patients and it reduces the dependency of the analysis on the operator \cite{Villa2018}.
	
\begin{figure}[t!] 
		\centering    
		\includegraphics[width=.85\textwidth]{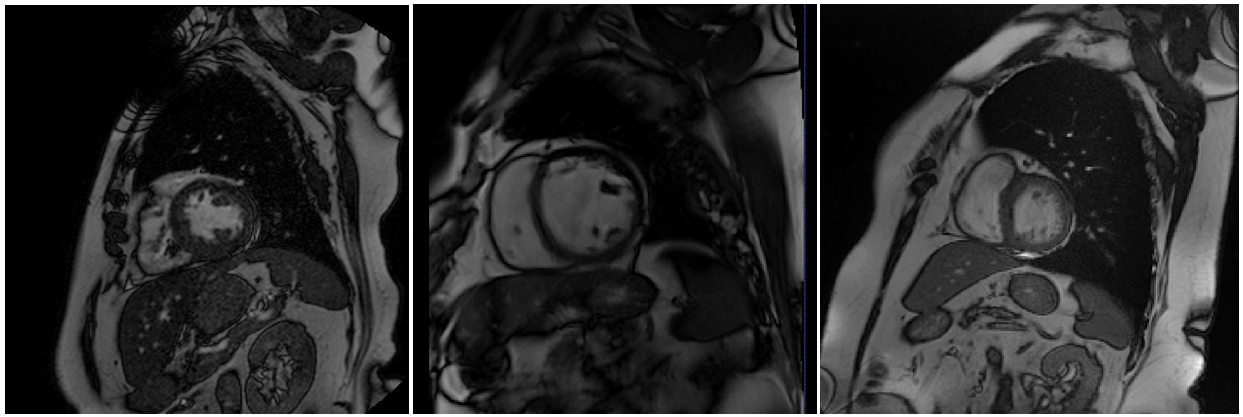}
		\caption{A comparison of images from different MR scanner vendors showing varying levels of signal, noise and contrast.}
		\label{fig:domains}
	\end{figure}

	However, the aforementioned studies are all limited by the homogeneity of the data that was used. They were either single/few centre studies or they used publicly available databases such as the UK Biobank which have uniform imaging protocols and few pathological cases. As such, the datasets do not match the variability seen in clinical practice and although the studies report good performance on their data, it is not guaranteed that they will generalise well to real-world settings. Images acquired using different scanners can have widely-varying levels of signal, noise, and contrast. Images acquired at different centres may be planned differently resulting in differing locations of the heart in the images, and distinct cardiovascular diseases can alter the shape of the heart. Some of these variations are shown in Fig.~\ref{fig:domains}. These variations introduce a so-called \textit{domain shift}, a change between the distribution of training data and the distribution of the data that the model is being applied to \cite{Dou2018}.
	
	Since deep learning models are known to not generalise well in the presence of such a domain shift \cite{Dou2018}, this leads to the need for techniques for domain adaption. The most simple technique for domain adaption is use of data augmentation to try to account for the variability between domains during training \cite{Chen2020}. It is, however, difficult to account for all possible variations in this manner. Other approaches include fine-tuning previously learned weights on a new domain and, more recently, conditional generative models, such as cycleGAN \cite{Yan2020}, have been used to translate images from a new domain to the domain of the model. However, training data would be required from the new domain for these approaches and thus they cannot be applied to completely unseen domains.
	
	Ganin et al. \cite{Ganin2016} proposed an approach to the domain-adversarial training of neural networks based on the assumption that for a model to generalise well to new domains the features learned by the model must be unable to discriminate between domains. This approach has been recently applied to domain adaption in histological images and shows promising results \cite{Lafarge2019}. In this work, domain-adversarial training is used for cine CMR segmentation and is shown to learn features that are domain invariant and generalise well to new unseen domains.

	\section{Materials and Methods}
	\subsection{Dataset}
	The dataset for this study was provided by the Multi-Centre, Multi-Vendor \& Multi-Disease (M\&Ms) Cardiac Image Segmentation Challenge\footnote[1]{\url{https://www.ub.edu/mnms/}} \cite{Campello}. The datasets consisted of 350 patients with a mix of healthy controls and patients with hypertrophic and dilated cardiomyopathies. It was acquired at six different clinical centres using MRI scanners from four different vendors. From this dataset 150 annotated patients from two different MRI vendors were made available for training as well as 25 patients from a third MRI vendor without annotations. There were 40 patients, 10 from each vendor, for validation and 160 patients, 40 from each vendor, for testing. These images were not available but results could be obtained from the challenge organisers. Since the test set for the challenge could only be used once, the comparison of our approach to the baseline is performed on this challenge validation set and 32 patients from the training set were used for internal validation. Expert annotations are provided for the left (LV) and right ventricle (RV) blood pools, as well as for the left ventricular myocardium (MYO).
	
	\subsection{Domain-Adversarial Learning}
\begin{figure}[h!] 
		\centering    
		\includegraphics[trim=3cm 0cm 9cm 0cm,clip=true, width=.9\textwidth]{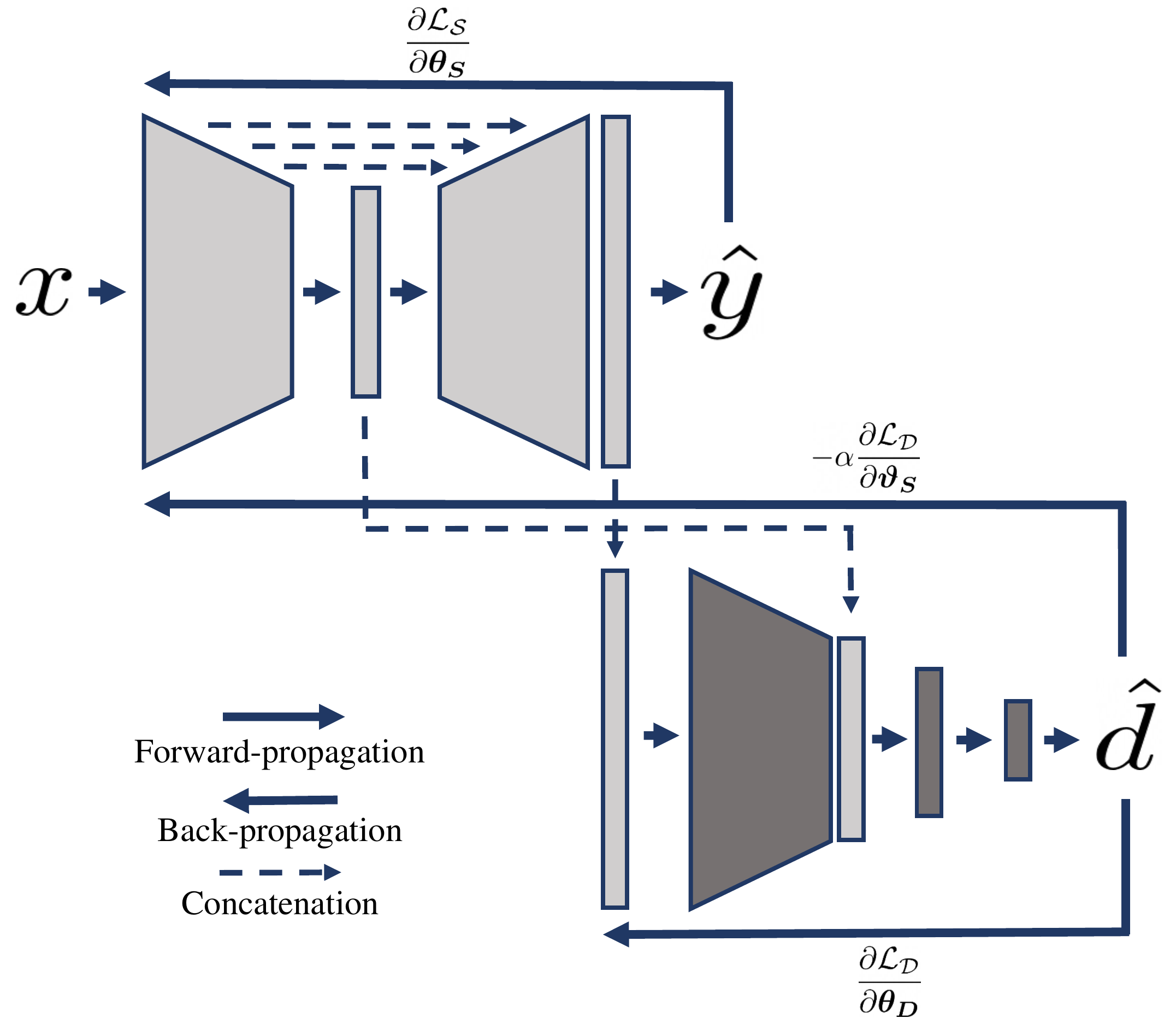}
		\caption{The flowchart of the domain-adversarial training pipeline. On top, the segmentation network is updated to minimise the segmentation loss. The intermediate representations of this network are then feed to a classification network which is optimised to discriminate between input domains. Finally, the adversarial back-propagation is used to update the parameters of the segmentation network to encourage domain invariance.}
		\label{fig:pipe}
	\end{figure}

	Domain-adversarial learning attempts to improve the generalisation of neural networks by encouraging them to learn features that do not depend on the domain of the input. In this situation, as well as learning a network for the segmentation task, we concurrently learn a classifier which attempts to discriminate between the input domains using the activations of the segmentation network. An adversarial training step is then introduced which is a third training step after the optimisation of the segmentation network and the optimisation of the domain discriminator. This adversarial step uses a gradient reversal to update the weights of the segmentation network to maximise the loss of the domain discriminator and thus prevents the domain information being recovered from the learned representations of the segmentation network. The three optimisation steps, using a standard (stochastic) gradient descent algorithm would be: 
	\begin{equation}
	\bm{\theta_S}  \leftarrow  \bm{\theta_S} - \lambda_S \frac{\partial \mathcal{L_S}}{\partial \bm{\theta_S}} \\
	\end{equation}
	\begin{equation}
	\bm{\theta_D}  \leftarrow  \bm{\theta_D} - \lambda_D \frac{\partial \mathcal{L_D}}{\partial \bm{\theta_D}}
	\end{equation}
	\begin{equation}
	\bm{\vartheta_S}  \leftarrow  \bm{\vartheta_S} + \alpha \lambda_S \frac{\partial \mathcal{L_D}}{\partial \bm{\vartheta_S}}
	\end{equation}
	where $\bm{\theta_S}, \lambda_S, \& \ \mathcal{L_S}$ and $\bm{\theta_D}, \lambda_D, \& \ \mathcal{L_D}$ are the parameters, learning rate, and loss function of the segmentation network and domain discriminator, respectively. $\bm{\vartheta_S} \subset \bm{\theta_S}$ are the parameters of the convolutional layers of the segmentation network and $\alpha \in [0,1]$ controls the strength of the adversarial update. The proposed training pipeline is illustrated in Fig.~\ref{fig:pipe}.
	
	\subsection{Implementation Details}
	Our TensorFlow implementation and trained model weights are provided on github\footnote[2]{\url{https://github.com/cianmscannell/da_cmr}}.
	
	The baseline model used in this work was a 2D U-Net model \cite{Ronneberger2015}. The model consists of four down-sampling blocks followed by max-pooling with each down-sampling block being made of up two convolution blocks containing a convolutional layer, batch normalisation, and ReLU activation. 
	
	The domain discriminator takes as input the activations from the penultimate and lowest resolution layers of the U-Net. The model consists of 4 convolutional layers with ReLU activations, batch normalisation, and max-pooling and three fully-connected layers also with ReLU activation. For the purpose of this work, different MRI scanner vendors were considered as distinct domains.
	
	Images are resampled to a uniform resolution of $1.25 \times 1.25$ \si{\milli\metre} and the network is trained on images cropped to $192 \times 192$ pixels. Pixel intensity values are normalised to the range $[0,1]$ before data augmentation. In this work, we use extensive spatial augmentation consisting of stochastic scaling, translation, rotation, and B-spline deformations using \textit{gryds} \cite{Eppenhof2019}. Intensity augmentation is performed by adding Gaussian noise and random intensity shifts. The intensities are further normalised using contrast limited adaptive histogram equalisation.
	
	The segmentation network is trained with batch size of 16 to optimise the sum of (1 - DSC) and the cross-entropy, where DSC is the Dice similarity coefficient. The domain discriminator is trained with a batch size of 20 and the cross-entropy loss. Optimisation is performed using the ADAM optimiser in both cases. Early-stopping is used on the domain classification accuracy to chose a model that has the least dependence on domain knowledge after the segmentation accuracy has plateaued.
	
	Images from the two annotated domains are used to train the segmentation networks while images from all three available domains are used to train the domain discriminator and compute the adversarial update, including the unlabelled images, as no ground-truth segmentations are required. Since the domain-adversarial training is unstable \cite{Lafarge2019}, the models are trained in stages. For the first 150 epochs, the segmentation U-Net is trained only with a learning rate of 1e-4. For the next 150 epochs, the domain discriminator is trained only with a learning rate of 1e-3. This initial training without the segmentation network updating allows it to achieve a baseline level of performance without its inputs (the activations of the segmentation network) changing. After this, both networks are trained together, including the adversarial update. The weighting of the adversarial update increases linearly from 0 to 1 over the next 150 epochs. The learning rate for the segmentation network (1e-3) is higher than the learning rate for the domain discriminator (1e-4) to allow the segmentation network to learn more quickly than the domain discriminator. At test time, predictions are made in a sliding window fashion over a 3x3 grid of patches with input size (192x192).
	
	This is compared to a baseline model trained in the exact same way, using the same intensity and data augmentation, except for the domain-adversarial update. The comparison was made using the nonparametric Mann-Whitney U test in SciPy \cite{Virtanen2020}.
	
	\section{Results}
\begin{figure}[htp!] 
		\centering    
		\includegraphics[width=.75\textwidth]{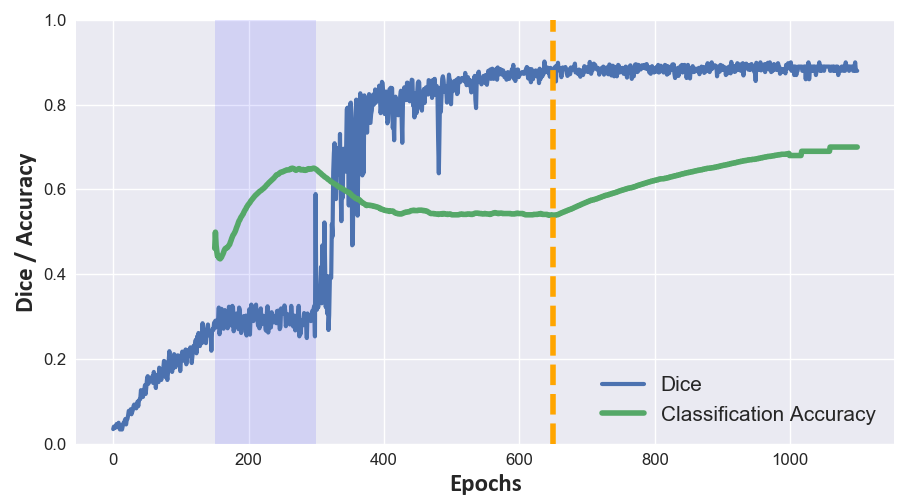}
		\caption{The loss curve, depicting the process of the domain-adversarial training. For the first 150 epochs, the segmentation network is trained only, then the domain discriminator is trained only (dark blue shaded area). Finally, both networks are trained together including adversarial updates. The orange dotted line indicates the network chosen using early-stopping.}
		\label{fig:loss}
	\end{figure}

\begin{figure}[h!]
		\centering    
		\includegraphics[width=1\textwidth]{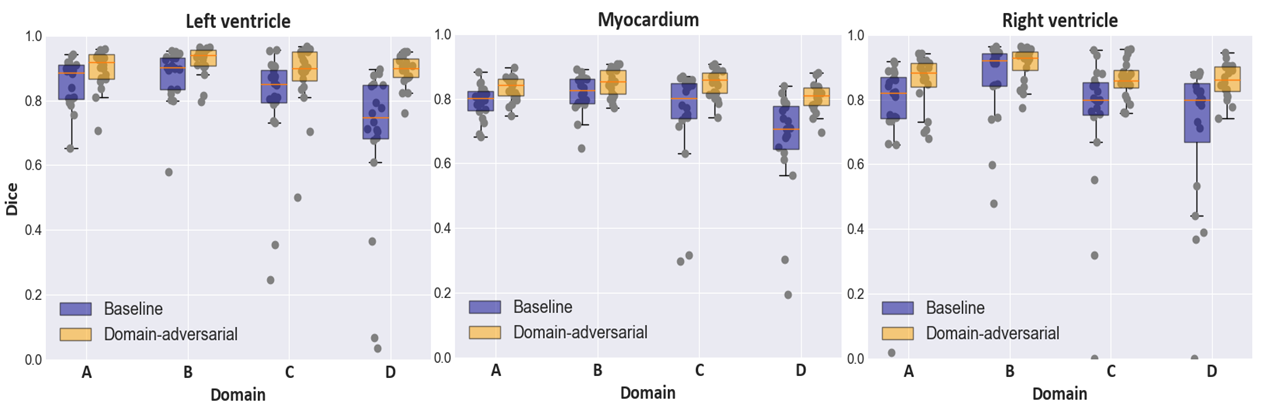}
		\caption{The two distributions of the Dice scores for the three classes for the different training domains. The domain-adversarial model has higher Dice with no drop-off in performance on the unseen domains (C and D). A drop-off in performance is seen from the training domains (A and B) to the unseen domains (C and D) for the baseline model.
			\newline}
		\label{fig:results}
	\end{figure}

	The loss and accuracy curves are shown for the domain adversarial training in Fig.~\ref{fig:loss}.
	The mean (SD) Dice similarity coefficient for the baseline model is 0.8 (0.18), 0.76 (0.13) and 0.77 (0.2) for the LV, MYO, and RV, respectively. The equivalent scores for the domain-adversarial model are 0.9 (0.07), 0.83 (0.05), and 0.87 (0.07). The Hausdorff distances for the three classes, in \si{\milli\metre}, are 17.6 (12.8), 26.3 (27.7), and 19.0 (12.1) for the baseline model and 12 (15.2), 17.4 (19.8), and 18.4 (33.4) for the domain-adversarial model. Figure~\ref{fig:results} further compares the baseline and domain-adversarial models showing the Dice for all three classes and how it varies across domains. The final test set results for our approach give Dice scores of 0.88 (0.1), 0.8 (0.09), and 0.84 (0.14) and Hausdorff distances (\si{\milli\metre}) of 14.54 (19.12), 17.36 (20.86), and 17.51 (19.01) for the LV, MYO, and RV, respectively. There is no performance drop-off across domains and a summary of all performance metrics is shown in the Table.~\ref{tab:results}
	
\begin{figure}[h!]
		\centering    
		\includegraphics[width=1\textwidth]{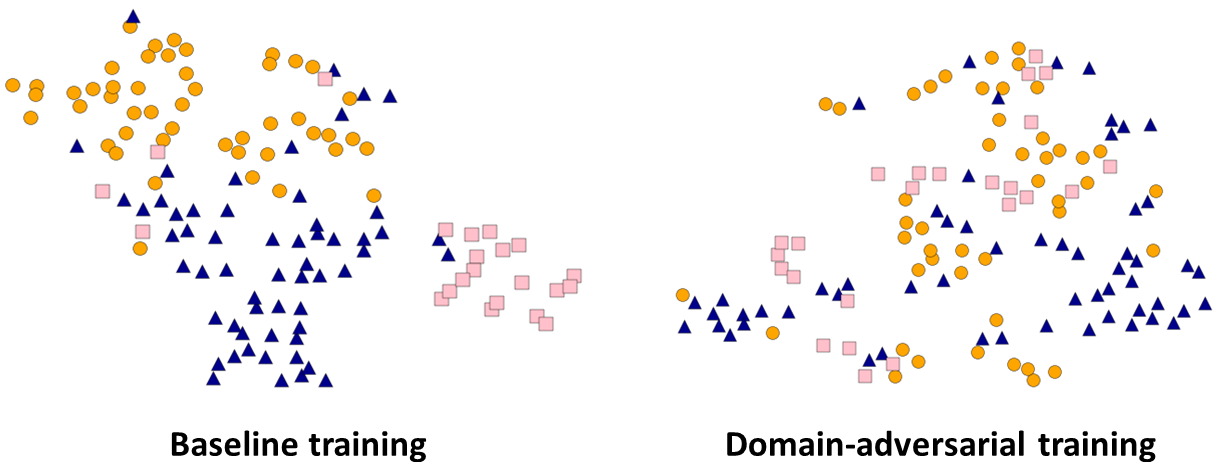}
		\caption{The t-SNE embeddings of the learned representations of 128 randomly selected image slices. The images are taken from the training domains (blue triangles and orange circles) and a previously unseen domain (pink squares). An image is represented as a vector by concatenating the means and standard deviations of the activations at the minimum resolution of the U-Net. The representations learned with baseline training (left) are compared to the representations learned in a domain-adversarial manner (right). This shows, qualitatively, that domain-adversarial training promotes the learning of features that are less dependent on the input domain.}
		\label{fig:tsne}
	\end{figure}

\begin{figure}
		\centering    
		\includegraphics[width=.85\textwidth]{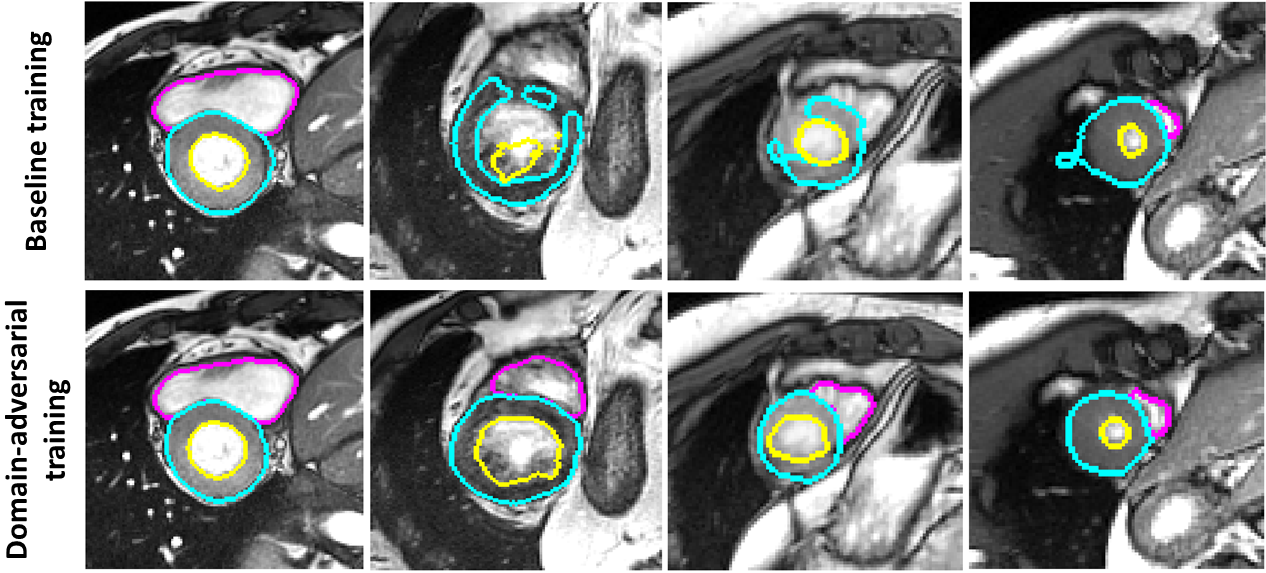}
		\caption{Example segmentations of images from a previously unseen domain, comparing standard training (top row) to domain-adversarial training (bottom row). Ground-truth segmentations are not available in this case but qualitatively superior performance is demonstrated with domain-adversarial training.}
		\label{fig:segmentations}
	\end{figure}

	The t-SNE embeddings of the learned representations of the baseline model are compared to the domain-adversarial learning approach in Fig.~\ref{fig:tsne}. With the baseline training there are clear clusters in the learned features indicating that they are domain dependent. For the domain-adversarial approach the learned features are more widely distributed in the embedding space, including for the domain not used for the training of the segmentation network (pink squares). Figure~\ref{fig:segmentations} shows the output segmentations for example images from one of the unseen domains, comparing the baseline model to the domain-adversarial approach.
	\newline

\begin{table} \label{tab:results}
	\caption{A summary, mean (standard deviation), of performance metrics across validation and test sets. For the validation set, a statistical comparison is further made (and p-value reported) for the Dice and Hausdorff (\si{\milli\metre}) metrics, respectively.}
	\centering
    	\makebox[\linewidth]{
	\begin{tabular}{cc|cc||cc|cc|cc}
		&       & \multicolumn{2}{c}{Test}   & \multicolumn{5}{c}{Validation}                                                                  \\ \hline
		&         & \multicolumn{2}{c}{Domain-adversarial} & \multicolumn{2}{c}{Baseline} &          \multicolumn{2}{c}{Domain-adversarial}  &  \\ \hline
		      Domain        &  & Dice & Hausdorff & Dice        & Hausdorff & Dice & Hausdorff                  & \multicolumn{2}{c}{p-value}            \\ \hline
		\multirow{3}{*}{A}   & LV      & 0.89 (0.08) & 16.78 (21.03)  &   0.85 (0.07)   &        15.23 (5.89)          &   0.9 (0.06)       &  14.5 (11.5) &      0.02     &  0.05   \\
		& MYO     & 0.81 (0.06) & 19.51 (22.18)  & 0.79 (0.05)       &      23.22 (21.19)                 &  0.83 (0.04)         &             17.37 (8.3)                           &      0.01          &   0.2 \\
		& RV      & 0.86 (0.08) & 16.86 (14.9)   &  0.77 (0.19)    &         21.55 (8.53)                        & 0.86 (0.08)          &      16.55 (6.05)                                  &       0.03          &  0.06 \\ \hline
		\multirow{3}{*}{B}   & LV      & 0.89 (0.01) & 11.65 (11.97)  & 0.88 (0.08)      &   10.7 (3.9)                              & 0.95 (0.05)         &     6.35 (2.45)                                   &  0.01                &  0.02    \\
		& MYO     & 0.82 (0.08) & 13.94 (12.16)  &  0.79 (0.04)     &        17.88 (25.77)                       &     0.82 (0.06)       &        9.53 (3.1)                                &          0.07          & 0.01  \\
		& RV      & 0.84 (0.16) & 17.5 (19.06)   & 0.86 (0.13)     &              11.36 (6.08)                   &  0.91 (0.05)         &                  9.98 (4.92)                      &         0.2       &  0.47  \\ \hline
		\multirow{3}{*}{C}   & LV      & 0.85 (0.13) & 17.39 (22.4)   & 0.8 (0.18)      &        18.07 (9.5)                        & 0.88 (0.11)         &        18.18 (26.37)                                &      0.02      &  $<$ 0.01     \\
		& MYO     & 0.75 (0.13) & 21.28 (26.07)  & 0.75 (0.16)      &             29.55 (28.01)                    &  0.84 (0.04)        &       24.96  (29.29)                               &        $<$ 0.01     &    0.06    \\
		& RV      & 0.8 (0.2)   & 17.29 (18.58)  &  0.74 (0.22)    &        17.29 (9.21)                        &   0.86 (0.06)       &     12.16 (4.56)                                   &    0.01      &   0.03       \\ \hline
		\multirow{3}{*}{D}   & LV      & 0.88 (0.05) & 12.16 (18.56)  & 0.68 (0.24)     &    26.5 (19.49)                             & 0.89 (0.05)         &        9.12 (3.45)                                &   $<$ 0.01      &    $<$ 0.01     \\
		& MYO     & 0.8 (0.05)  & 14.69 (19.53)  &     0.67 (0.16) &          34.35 (32)                       &  0.8 (0.04)         &            17.79 (22.72)                            &   $<$ 0.01      &    0.01     \\
		& RV      & 0.87 (0.11) & 18.31 (22.7)   &  0.7 (0.23)    &         25.88 (16.52)                        &  0.86 (0.05)         &             34.89 (63.21)                           &    $<$ 0.01     &    0.02     \\ \hline
		\multirow{3}{*}{All} & LV      & 0.88 (0.1)  & 14.54 (19.12)  &    0.8 (0.18)  &             17.6  (12.8)                  &    0.9 (0.07)      &                       12 (15.2)                 &    $<$ 0.01    &  $<$ 0.01     \\
		& MYO     & 0.8 (0.09)  & 17.36 (20.86)  &   0.76 (0.13)   &              26.3 (27.7)                   &   0.83 (0.05)       &                 17.4 (19.8)                       &     $<$ 0.01        &   $<$ 0.01    \\
		& RV      & 0.84 (0.14) & 17.51 (19.01)  &   0.77 (0.2)    &              19 (12.1)                   &   0.87 (0.07)       &                      18.4 (33.4)                  &      $<$ 0.01     &  $<$ 0.01     
	\end{tabular}
}
\end{table}

	\section{Discussion}
	This paper deals with the challenge of the generalisation of deep learning models across different data domains. In particular, it deals with the segmentation of short-axis cine CMR images which vary significantly across MRI scanner vendors. 
	
	We have demonstrated that domain-adversarial learning can improve the generalisation of a U-Net model trained to segment short-axis cine CMR images. We trained the model using annotated data from two vendors and incorporated the domain information using unannotated data from a third vendor. The model can be applied on data from unseen domains (i.e from different MRI vendors) with no significant performance decrease. Although baseline training includes intensity normalisation and contrast normalisation through the use of histogram equilisation, lower performance is found on data from unseen domains. The comparison over the whole validation set shows statistically significant improvements using domain-adversarial learning, for the segmentation of all structures. This indicates that there are further variations across vendors, in addition to intensity and contrast, that the network relies on for the segmentation. The improved performance of the domain-adversarial approach shows that it can avoid over-fitting to this domain information and that it can learn more general, domain-invariant features. The performance of the baseline model is also worse on the training domains. This is likely because even data from the same vendor can be from different clinical centres which in itself may constitute a domain shift and because the domain-adversarial update has a regularising effect on training.
	
	Domain-adversarial learning has the benefit, over alternative domain-adaption strategies, of not requiring any data from the domains that the model is being applied on and not requiring any re-training. It also allows the incorporation of data without annotations in the domain discriminator and adversarial update so that unlabelled data can still be used to encourage domain invariance.
	
	As a limitation, the domain-adversarial training of neural networks is an unstable process. It relies on the ad-hoc hyper-parameter tuning more than convention training and work to address this would be needed for the technique to be used more commonly. Though the dataset used does include some pathological cases, these are not with a wide range of conditions and it remains to be determined if the domain-adversarial approach can still adapt to patients with different conditions that alter the geometry of the heart, such as ischaemic heart disease.
	
	In conclusion, domain-adversarial learning improves the ability of deep learning approaches for CMR image segmentation to generalise to unseen data from new domains. It does not require data from the new domains or to fine-tune/adapt the network. 
	
	\section{Acknowledgements}
	The authors of this paper declare that the segmentation method they implemented for participation in the M\&Ms challenge has not used any pre-trained models nor additional MRI datasets other than those provided by the organizers. 
	
	The authors acknowledge financial support from the King’s College London \& Imperial College London EPSRC Centre for Doctoral Training in Medical Imaging (EP/L015226/1); Philips Healthcare; and The Centre of Excellence in Medical Engineering funded by the Wellcome Trust and EPSRC under grant number WT 203148/Z/16/Z.
	

\end{document}